\documentclass[twocolumn,superscriptaddress,pra]{revtex4-1}
\pdfoutput=1
\usepackage{mathrsfs,braket}
\usepackage{amssymb, amsbsy, amsmath, latexsym, dsfont, array, layout,
graphicx,mathrsfs,color,ulem,bm}

\newcommand{\one}{\mathds{1}}

\begin{document}

\title{Observation of emergent momentum-time skyrmions in\\ parity-time-symmetric non-unitary quench dynamics}
\author{Kunkun Wang}
\affiliation{Beijing Computational Science Research Center, Beijing 100084, China}
\affiliation{Department of Physics, Southeast University, Nanjing 211189, China}
\author{Xingze Qiu}
\affiliation{Key Laboratory of Quantum Information, University of Science and Technology of China, CAS, Hefei 230026, China}
\affiliation{CAS Center For Excellence in Quantum Information and Quantum Physics}
\author{Lei Xiao}
\affiliation{Beijing Computational Science Research Center, Beijing 100084, China}
\affiliation{Department of Physics, Southeast University, Nanjing 211189, China}
\author{Xiang Zhan}
\affiliation{Beijing Computational Science Research Center, Beijing 100084, China}
\affiliation{Department of Physics, Southeast University, Nanjing 211189, China}
\author{Zhihao Bian}
\affiliation{Beijing Computational Science Research Center, Beijing 100084, China}
\affiliation{Department of Physics, Southeast University, Nanjing 211189, China}
\author{Barry C. Sanders}
\affiliation{Institute for Quantum Science and Technology, University of Calgary, Alberta T2N 1N4, Canada}
\affiliation{Program in Quantum Information Science, Canadian Institute for Advanced Research, Toronto, Ontario M5G 1Z8, Canada}
\affiliation{Shanghai Branch, National Laboratory for Physical Sciences at Microscale, University of Science and Technology of China, Shanghai 201315, China}
\author{Wei Yi}\email{wyiz@ustc.edu.cn}
\affiliation{Key Laboratory of Quantum Information, University of Science and Technology of China, CAS, Hefei 230026, China}
\affiliation{CAS Center For Excellence in Quantum Information and Quantum Physics}
\author{Peng Xue}\email{gnep.eux@gmail.com}
\affiliation{Beijing Computational Science Research Center, Beijing 100084, China}
\affiliation{Department of Physics, Southeast University, Nanjing 211189, China}
\affiliation{State Key Laboratory of Precision Spectroscopy, East China Normal University, Shanghai 200062, China}

\begin{abstract}
Topology in quench dynamics gives rise to intriguing dynamic topological phenomena, which are intimately connected to the topology of static Hamiltonians yet challenging to probe experimentally.
Here we experimentally detect momentum-time skyrmions in parity-time ($\mathcal{PT}$)-symmetric non-unitary quench dynamics, which are protected by dynamic Chern numbers defined for the emergent momentum-time manifold. Specifically, we experimentally simulate non-unitary quench dynamics of $\mathcal{PT}$-symmetric topological systems using single-photon discrete-time quantum walks, and
demonstrate emergence of skyrmions by constructing the time-dependent non-Hermitian density matrix via direct measurements in position space.
Our work experimentally reveals the interplay of $\mathcal{PT}$ symmetry and quench dynamics in inducing emergent topological structures,
and highlights the application of discrete-time quantum walks for the study of dynamic topological phenomena.
\end{abstract}

\maketitle


\section{introduction}

Topological phases feature a wealth of fascinating properties governed by the geometry of their ground-state wave functions at equilibrium~\cite{HKrmp10,QZrmp11},
but topological phenomena also manifest as non-equilibrium quantum dynamics in driven-dissipative~\cite{Zollerdiss} and Floquet systems~\cite{Levin13,Lindner16,Vishwanath16,rigol}, as well as in quench processes~\cite{Bhaseen15,Budich16,Refael16,Zhai17,Chen17,Ueda17,wytheory18,Xiong-Jun}. The experimental detection of these dynamic topological phenomena is challenging since it requires full control and access of the time-evolved state. In recent experiments with ultracold atoms, topological objects such as vortices, links and rings have been identified in the quench dynamics of topological systems via time- and momentum-resolved tomography~\cite{Weitenberg17,Weitenberg1709,shuaichen}.
Here we experimentally establish discrete-time photonic quantum walks (QWs) as another promising arena for engineering and detecting dynamic topological phenomena. Compared to cold atomic gases, the relative ease of introducing loss in photonic systems further enables us to experimentally investigate novel dynamic topological phenomena in the non-unitary regime, where parity-time ($\mathcal{PT}$) symmetry plays an important role.

In discrete-time photonic QWs~\cite{CSPRA,Cardano2016,KB+12,Cardano2017,BNE+17,WXQ+18}, single photons, starting from their initial states, are subject to repeated unitary operations~\cite{PLM+10}. While QW dynamics support Floquet topological phases (FTPs)~\cite{KB+12,Cardano2017,BNE+17,WXQ+18,PTsymm2,pxprl}, discrete-time QWs can also be viewed as stroboscopic simulation of quench dynamics between FTPs, during which dynamic topological phenomena should occur. However, the enticing possibility of QWs in unveiling dynamic topological phenomena in quench processes has not been explored.


We bridge this gap by experimentally detecting dynamic skyrmion structures in $\mathcal{PT}$-symmetric one-dimensional QWs of single photons. Originally proposed in high-energy physics~\cite{sky1} and later experimentally observed in magnetic and optical configurations~\cite{magsky,crysky,optsky}, skyrmions are a type of topologically stable defects featuring a three-component vector field in two dimensions. In QW dynamics, dynamic skyrmions manifest themselves in the momentum-time spin texture of the time-evolved density matrix, and are protected by quantized dynamic Chern numbers in emergent momentum-time submanifolds~\cite{Chen17,Ueda17,wytheory18}.
We apply projective and interference-based measurements in position space for the construction of time-dependent density matrix, rather than the time-resolved tomography. Such a practice allows for direct measurements of the density matrix at each time step, which significantly reduces the systematic error introduced by the least-square algorithm in tomographic measurements.

We confirm the emergence of dynamic skyrmion structures when QW dynamics correspond to quenches between distinct FTPs in the $\mathcal{PT}$-symmetry-unbroken regime, where the dynamics is coherent despite being non-unitary. Effective coherent dynamics is manifested as temporal oscillatory behavior inherent in off-diagonal density-matrix elements. Such oscillatory phenomena reflect the system's ability to fully retrieve information temporarily lost to the environment by PT dynamics in the unbroken-symmetry regime~\cite{Uedainfo}.
By contrast, when the system is quenched into the $\mathcal{PT}$-symmetry-broken regime, skyrmions are absent in the momentum-time space, as the dynamics become incoherent.
Our work unveils the fascinating relation between emergent topology and $\mathcal{PT}$-symmetric non-unitary dynamics, and is the first experiment to showcase the prowess of QWs in revealing dynamic topological structures and invariants in quench dynamics.

\begin{figure*}
\includegraphics[width=0.9\textwidth]{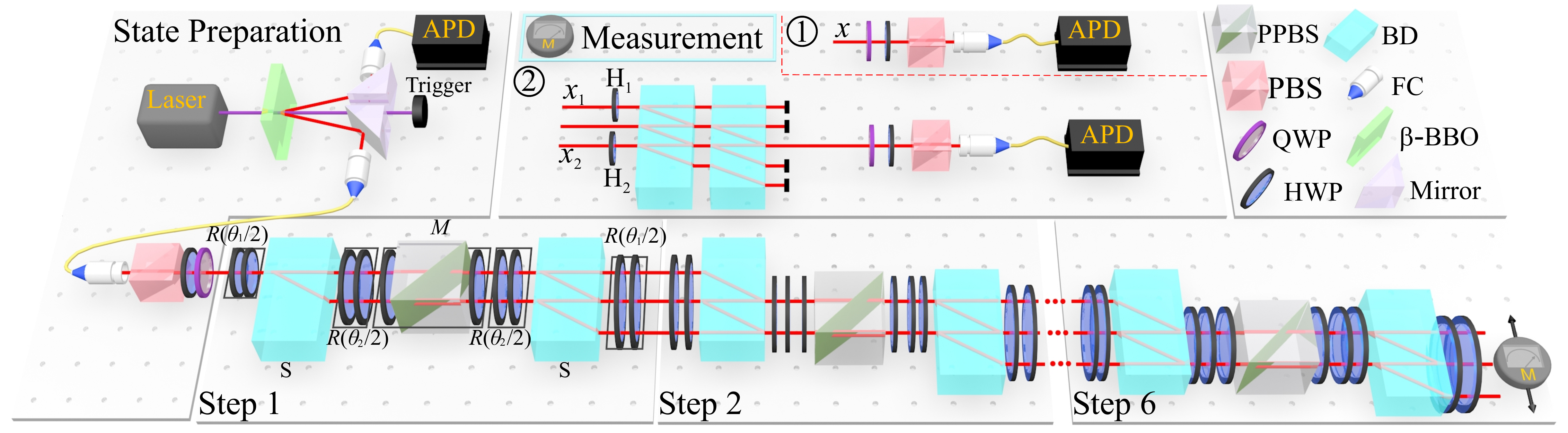}
\caption{Experimental setup for detecting momentum-time skyrmions in non-unitary QWs. Photons are generated via spontaneous parametric down
conversion through a Type-I non-linear $\beta$-Barium-Borate (BBO) crystal. The single signal photon is heralded by the corresponding trigger photon and can be prepared in an arbitrary linear polarization state via a polarizing beam splitter (PBS) and wave plates. Conditional shift operation $S$ and coin rotation $R$ are realized by a beam displacer (BD) and two half-wave plates (HWPs), respectively. For non-unitary QWs, a sandwich-type HWP-PPBS-HWP setup is inserted to introduce non-unitarity, where PPBS is an abbreviation for partially polarizing beam splitters. Two kinds of measurements, including projective measurements and interference-based measurements, are applied before the signal and heralding photons are detected by avalanche photodiodes (APDs).
}
\label{fig:setup}
\end{figure*}


\section{Quench dynamics in $\mathcal{PT}$-symmetric QWs}

We experimentally implement $\mathcal{PT}$-symmetric non-unitary QWs on a one-dimensional lattice $L$ ($L\in \mathbb{Z}$) with single photons in the cascaded interferometric network illustrated in Fig.~\ref{fig:setup}. The corresponding Floquet operator is
\begin{align}
U=R\left(\frac{\theta_1}{2}\right)SR\left(\frac{\theta_2}{2}\right)MR\left(\frac{\theta_2}{2}\right)SR\left(\frac{\theta_1}{2}\right),
\label{eq:defU}
\end{align}
where $R(\theta)$ rotates coin states (encoded in the horizontal and vertical polarizations of single photons $\ket{H}$ and $\ket{V}$) by $\theta$ about the $y$-axis, and $S$ moves the photon to neighbouring spatial modes depending on its polarization (see Appendix).
The loss operator $M=\one_\text{w}\otimes\left(\ket{+}\bra{+}+\sqrt{1-p}\ket{-}\bra{-}\right)$
enforces a partial measurement in the basis $\ket{\pm}=\left(\ket{H}\pm\ket{V}\right)/\sqrt{2}$ at each time step with a success probability $p\in[0,1]$. Here $\one_\text{w}=\sum_x\ket{x}\bra{x}$ with $\ket{x}$ ($x\in L$) denoting the spatial mode.
Note that the non-unitary QW driven by $U$ reduces to a unitary one $p=0$.

\begin{figure}
\includegraphics[width=0.5\textwidth]{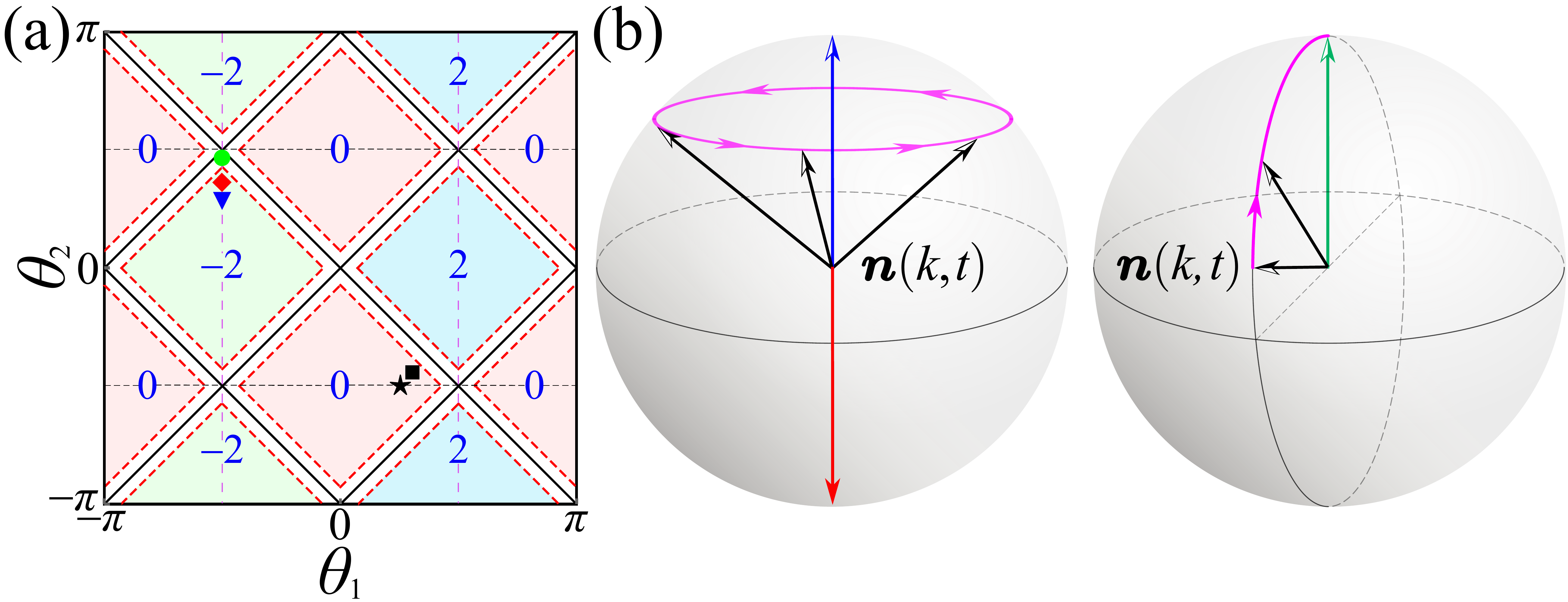}
\caption{Phase diagram and schematic illustrations of non-unitary QW dynamics. (a) Phase diagram for QWs governed by the Floquet operator $U$ in Eq.~(\ref{eq:defU}), with the corresponding topological numbers $\nu$ as a function of coin parameters $(\theta_1,\theta_2)$. Solid black lines are the topological phase boundary, dashed red lines represent boundaries between $\mathcal{PT}$-symmetry-unbroken and
broken regimes. Black star represents coin parameters of $U^\text{i}$, of which the initial state $|\psi^\text{i}\rangle$ is an eigenstate.
Red diamond, blue triangle, and green circle correspond to coin parameters of the final Floquet operator $U^\text{f}$ in Fig.~\ref{fig:oscillation}(a), Fig.~\ref{fig:oscillation}(b), and Fig.~\ref{fig:broken}, respectively. The black square correspond to coin parameters of $U^\text{f}$ in Fig.~6. (b) Schematic illustrations of the time evolution of $\bm{n}(k,t)$ on a Bloch sphere when $E^\text{f}_k$ is real (left) and imaginary (right), respectively. Blue and red arrows point to fixed points. The green arrow indicates steady state in the long-time limit. Black arrows indicate the direction of $\bm{n}(k,t)$ at different times.}
\label{fig:bloch}
\end{figure}

QWs governed by $U$ stroboscopically simulate non-unitary time evolutions driven by the effective Hamiltonian $H_{\text{eff}}$, with $U=e^{-i H_{\text{eff}}}$. We define the quasienergy $\epsilon$ and eigenstate $|\psi\rangle$ as
$U|\psi\rangle=\gamma^{-1} e^{-i \epsilon}|\psi\rangle$, where $\gamma=(1-p)^{-\frac{1}{4}}$.
$U$ possesses passive $\mathcal{PT}$ symmetry with $\mathcal{PT} \gamma U \left(\mathcal{PT}\right)^{-1}=\gamma^{-1}U^{-1}$, where $\mathcal{PT}=\sum_x|-x\rangle\langle x|\otimes\sigma_3\mathcal{K}$, $\sigma_3=|H\rangle\langle H|-|V\rangle\langle V|$, and $\mathcal{K}$ is the complex conjugation.
It follows that $\epsilon$ is entirely real in the $\mathcal{PT}$-symmetry-unbroken regime, and can take imaginary values in regimes when $\mathcal{PT}$ symmetry is spontaneously broken~\cite{BB98,BBJ02,B07,pxdqpt}. $U$ also features topological properties, characterized by winding numbers defined through the global Berry phase~\cite{GW88,LH13,Lieu18}. We show the topological phase diagram of the system in Fig.~\ref{fig:bloch}(a), where distinct FTPs are marked by their corresponding winding numbers. The boundaries between $\mathcal{PT}$-symmetry-unbroken and -broken regimes are also shown in red-dashed lines, with $\mathcal{PT}$-symmetry-broken regimes surrounding topological phase boundaries.

To simulate quench dynamics, we initialize the walker photon in the eigenstate $|\psi^\text{i}\rangle$ of a Floquet operator $U^\text{i}=\text{e}^{-iH^\text{i}_{\text{eff}}}$, characterized by coin parameters $(\theta^\text{i}_1,\theta^\text{i}_2)$. The walker at the $t$-th time step is given by $|\psi(t)\rangle=e^{-iH_{\text{eff}}t}|\psi^\text{i}\rangle$, such that the resulting QW can be identified as a sudden quench between $H^\text{i}_\text{eff}$ and $H_\text{eff}$. Adopting notations in typical quench dynamics, we denote $U$ and $H_\text{eff}$ as $U^\text{f}$ and $H^\text{f}_\text{eff}$ in the following, characterized by coin parameters $(\theta^\text{f}_1,\theta^\text{f}_2)$.

\begin{figure*}
\includegraphics[width=\textwidth]{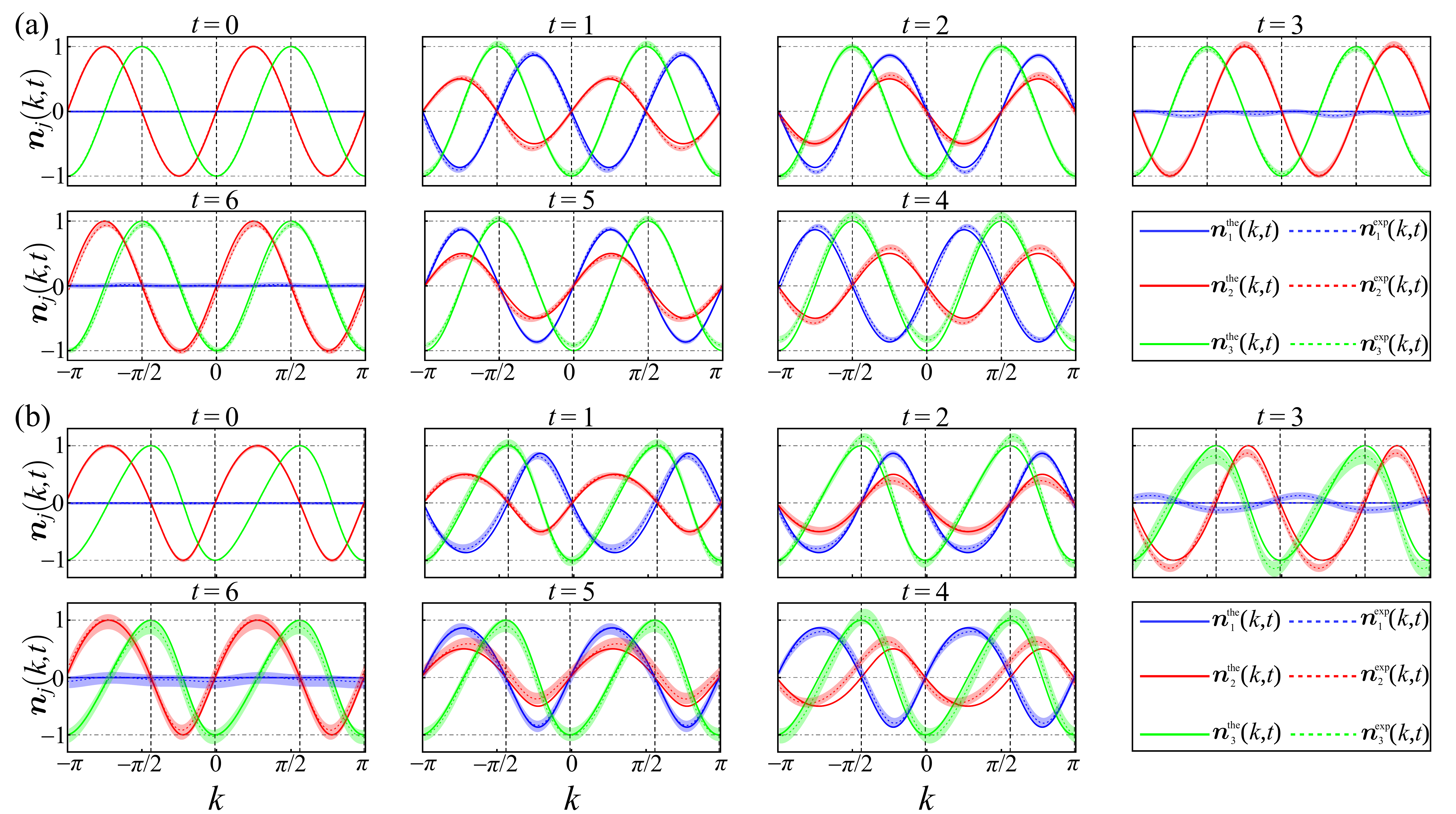}
\caption{Experimental results of $\bm{n}(k,t)$. Time-evolution of $\bm{n}(k,t)$ up to $t=6$ for quench processes between (a) an initial unitary Floquet operator characterized by $(\theta^\text{i}_1=\pi/4,\theta^\text{i}_2=-\pi/2)$ and a final unitary Floquet operator characterized by $(\theta^\text{f}_1=-\pi/2,\theta^\text{f}_2=\pi/3)$; and (b) an initial non-unitary Floquet operator characterized by $(\theta^\text{i}_1=\pi/4,\theta^\text{i}_2=-\pi/2)$ and a final non-unitary Floquet operator characterized by $(\theta^\text{f}_1=-\pi/2,\theta^\text{f}_2=\arcsin(\frac{1}{\alpha}\cos\frac{\pi}{6}))$. The period of oscillations is $t_0=6$ for all $k$. Fixed points are located at $\{-\pi,-\pi/2,0,\pi/2\}$ for unitary dynamics (a), and at $\{-0.4399\pi,-0.0099\pi,0.5901\pi,0.9901\pi\}$ for non-unitary dynamics (b). Shadings indicate experimental error bars which are are due to photon-counting statistics.}
\label{fig:oscillation}
\end{figure*}

\section{Fixed points and emergent skyrmions}

Due to the lattice translational symmetry of $U^{\text{i},\text{f}}$, dynamics in different quasi-momentum~$k$-sectors are decoupled.
We consider the case where $U^\text{i}$ is in the $\mathcal{PT}$-symmetry-unbroken regime, with the initial state $|\psi_{k,-}^\text{i}\rangle$ in each $k$-sector satisfying $U^{\text{i}}_k|\psi_{k,-}^{\text{i}}\rangle=\gamma^{-1} e^{-i\epsilon^\text{i}_{k,-}}|\psi_{k,-}^{\text{i}}\rangle$.
Likewise, we have $U^{\text{f}}_k|\psi_{k,\pm}^{\text{f}}\rangle=\gamma^{-1} e^{- i\epsilon^\text{f}_{k,\pm}}|\psi_{k,\pm}^{\text{f}}\rangle$, where we denote quasienergies of $U^{\text{i},\text{f}}_k$ as $\epsilon^{\text{i},\text{f}}_{k,\pm}$, with $\epsilon^{\text{i},\text{f}}_{k,\pm}=\pm E^{\text{i},\text{f}}_k$.

By invoking the biorthogonal basis~\cite{DCB}, the non-unitary time evolution of the system is captured by a non-Hermitian density matrix, which can be written as~\cite{wytheory18}
\begin{align}
\rho(k,t)=\frac{1}{2}\left[\tau_0+\bm{n}(k,t)\cdot\bm{\tau}\right],
\label{Eq:dmbloch}
\end{align}
where $\bm{n}(k,t)=(n_1,n_2,n_3)$, $\bm{\tau}=(\tau_1,\tau_2,\tau_3)$, $\tau_i=\sum_{\mu,\nu=\pm}|\psi_{k,\mu}^{\text{f}}\rangle\sigma^{\mu\nu}_i\langle\chi_{k,\nu}^{\text{f}}|$ ($i=0,1,2,3$), and $\langle\chi^{\text{f}}_{k,\mu}|$ $\left(|\psi^{\text{f}}_{k,\mu}\rangle\right)$ is the left (right) eigenvector of $U^\text{f}_k$. Here,
$\sigma_0$ is a $2\times 2$ identity matrix, and $\sigma_i$ $(i=1,2,3)$ is the corresponding standard Pauli matrix.

A key advantage of adopting Eq.~(\ref{Eq:dmbloch}) is that $\bm{n}(k,t)$ becomes a real unit vector, which enables us to visualize the non-unitary dynamics on a Bloch sphere. As illustrated in Fig.~\ref{fig:bloch}(b), when $E_{k}^{\text{f}}$ is real, $\bm{n}(k,t)$ rotates around poles of the Bloch sphere with a period $t_0=\pi/E_{k}^{\text{f}}$. Thus, momenta corresponding to poles of the Bloch sphere are identified as two different kinds of fixed points, where the density matrices do not evolve in time.
In contrast, when $E_{k}^{\text{f}}$ is imaginary, there are no fixed points in the dynamics, as $\bm{n}(k,t)$ asymptotically approaches the north pole in the long-time limit [see Fig.~\ref{fig:bloch}(b)].

\begin{figure*}
\includegraphics[width=\textwidth]{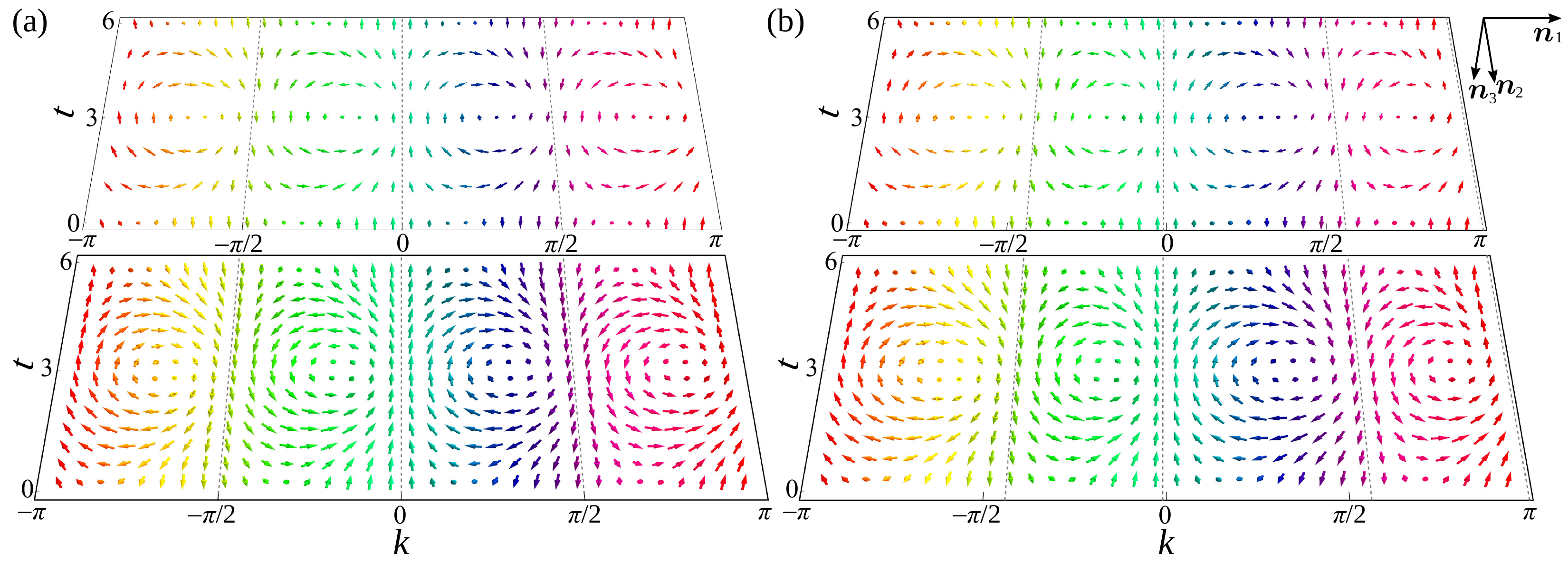}
\caption{Experimental results of spin texture $\bm {n}(k,t)$. Experimental (upper layer) and theoretical results (lower layer) of spin texture $\bm {n}(k,t)$ in the momentum-time space for quench
processes corresponding to (a) Fig.~\ref{fig:oscillation}(a), and (b) Fig.~\ref{fig:oscillation}(b), respectively. The temporal resolution of experimental measurements is limited by discrete time steps of QWs, whereas we adopt a better resolution in theoretical results for a clearer view of skyrmions.}
\label{fig:vector}
\end{figure*}

When $U^\text{i}$ and $U^\text{f}$ belong with distinct FTPs in the $\mathcal{PT}$-symmetry-unbroken regime, fixed points of different kinds necessarily emerge in pairs~\cite{wytheory18,pxdqpt}. Each momentum submanifold between a pair of distinct fixed points can be combined with the $S^1$ topology of the periodic time evolution to form an emergent $S^2$ momentum-time manifold, which can be mapped to the $S^2$ Bloch sphere of $\bm{n}(k,t)$. The Chern number characterizing such an $S^2\rightarrow S^2$ mapping is finite and gives rise to intriguing skyrmion structures in the emergent momentum-time manifolds.

To probe fixed points and momentum-time skyrmions, we perform projective and interference-based measurements to construct the Hermitian density matrix $\rho'(k,t)=|\psi_k(t)\rangle\langle\psi_k(t)|$, from which we calculate the non-Hermitian density matrix $\rho(k,t)$ and determine $\bm{n}(k,t)$ through $\bm{n}(k,t)=\text{Tr}\left[\rho(k,t)\cdot\bm{\tau}\right]$. We emphasize that whereas the Hermitian density matrix $\rho'(k,t)$ is experimentally accessible, it is difficult to visualize non-unitary dynamics on a Bloch sphere starting from it, and skyrmion structures would remain hidden in the dynamics.

\section{Dynamics in the $\mathcal{PT}$-symmetry-unbroken regime}

We first study fixed points and momentum-time skyrmions in the $\mathcal{PT}$-symmetry-unbroken regime. For comparison, we also experimentally characterize these quantities in unitary dynamics. We initialize the walker on a localized lattice site $|x=0\rangle$ and in the coin state $\ket{\psi^\text{i}_{-}}_\text{c}$. Here, $|x\rangle$ denotes the spatial mode. Importantly, $\ket{\psi^\text{i}_{k,-}}=\ket{\psi^\text{i}_{-}}_\text{c}$ is an eigenstate of $U^\text{i}_k$ for all $k$, with the corresponding $(\theta^\text{i}_1,\theta^\text{i}_2)$ on blue dashed lines in Fig.~2(a). Without loss of generality, we choose $(\theta^\text{i}_1=\pi/4,\theta^\text{i}_2=-\pi/2)$ for both the unitary and  non-unitary cases.

For the first case of study, we implement
unitary QWs with $\ket{\psi^\text{i}_{-}}_\text{c}=(\ket{H}+i\ket{V})/\sqrt{2}$ and $(\theta^\text{f}_1=-\pi/2,\theta^\text{f}_2=\pi/3)$, which simulate quench processes between FTPs with $\nu^\text{i}=0$ and $\nu^\text{f}=-2$. We have chosen $(\theta^\text{f}_1,\theta^\text{f}_2)$ on purple dashed lines,
where qusienergy bands are flat. Oscillatory dynamics of $\bm{n}(k,t)$ in different $k$-sectors thus feature the same period, as illustrated in Fig.~\ref{fig:oscillation}(a). We identify fixed points of unitary dynamics at high-symmetry points of the Brillioun zone $\{-\pi,-\pi/2,0,\pi/2\}$, where $\bm{n}(k,t)$ become independent of time.

For the second case of study, we implement non-unitary QWs with $p=0.36$, $\ket{\psi^\text{i}_{-}}_\text{c}=0.7606\ket{H}+0.6492i\ket{V}$, and $\left[\theta^\text{f}_1=-\pi/2,\theta^\text{f}_2=\arcsin(\frac{1}{\alpha}\cos\frac{\pi}{6})\right]$ (here $\alpha=\frac{\gamma}{2}(1+\sqrt{1-p})$). The post-quench FTP is in the $\mathcal{PT}$-symmetry unbroken regime with $\nu^\text{f}=-2$. As shown in Fig.~\ref{fig:oscillation}(b), dynamics of $\bm{n}(k,t)$ is still oscillatory, but fixed points are shifted away from the high-symmetry points, consistent with theoretical predictions.

In Fig.~\ref{fig:vector}, we plot $\bm{n}(k,t)$ in the momentum-time space. The oscillatory behavior in $\bm{n}(k,t)$ is then manifested as momentum-time skyrmions, which are protected by dynamic Chern numbers defined on the corresponding momentum-time submanifold.
By contrast, when the system is quenched between FTPs with the same winding number, skyrmion-lattice structures are no longer present (see Fig.~6 in the Appendix).

\section{Dynamics in the $\mathcal{PT}$-symmetry-broken regime}

\begin{figure*}
\includegraphics[width=\textwidth]{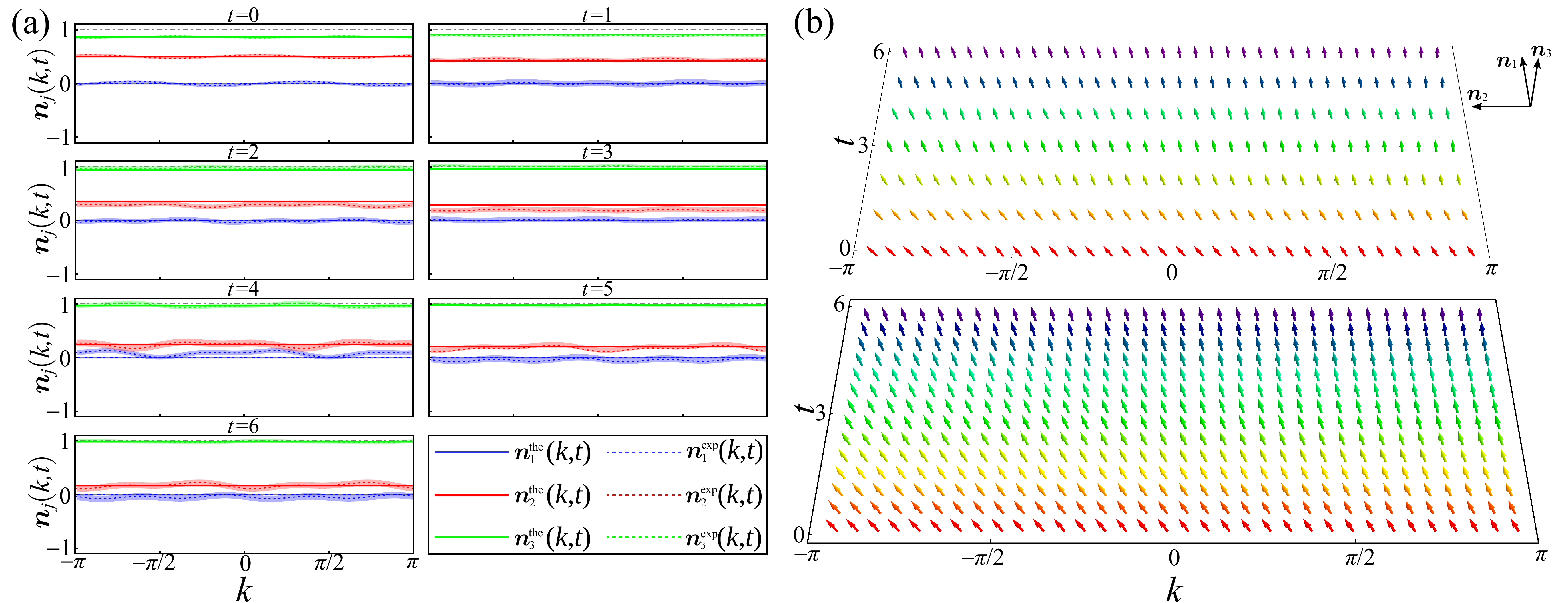}
\caption{Experimental results for the $\mathcal{PT}$-symmetric broken QW dynamics. (a) Time-evolution and (b) spin textures of $\bm{n}(k,t)$ in the momentum-time space for a quench process between the initial non-unitary Floquet operator given by $(\theta^\text{i}_1=\pi/4,\theta^\text{i}_2=-\pi/2)$ and the final $\mathcal{PT}$-symmetry-broken Floquet operator given by $\left[\theta^\text{f}_1=-\pi/2,\theta^\text{f}_2=\frac{1}{2}(\pi-\arccos\frac{1}{\alpha})\right]$.}
\label{fig:broken}
\end{figure*}

We now turn to the case where $U^\text{f}$ belong with the $\mathcal{PT}$-symmetry-broken regime. We initialize the walker on a localized lattice site in the coin state $\left(\ket{H}+\ket{V}\right)/\sqrt{2}$, and evolve it under $U^\text{f}$ characterized by $\left[\theta^\text{f}_1=-\pi/2,\theta^\text{f}_2=\frac{1}{2}(\pi-\arccos\frac{1}{\alpha})\right]$, which is in the $\mathcal{PT}$-symmetry-broken regime with $\nu^\text{f}=-2$. We note that for contrast, $U^\text{f}$ is chosen such that its quasienergy spectra are flat and completely imaginary.
As shown in Fig.~\ref{fig:broken}(a), there is no periodical evolution in $\bm{n}(k,t)$ anymore. Instead, different components of $\bm{n}(k,t)$ slowly approach a steady state with $\bm{n}=(0,0,1)$ in the long-time limit. This is more clearly seen in the momentum-time space in Fig.~\ref{fig:broken}(b), where skyrmion structures are absent and vectors in all $k$-sectors tend to point out of the plane in the long-time limit. We note that dynamics of $\bm{n}(k,t)$ here is insensitive to the choice of initial state, as the system always relaxes to the steady state at long times.

\section{Discussion}

By simulating quench dynamics of topological systems using photonic QWs, we have revealed emergent momentum-time skyrmions, protected by dynamic Chern numbers defined on the momentum-time submanifolds. These emergent topological phenomena are underpinned by fixed points of dynamics, which can exist for both unitary and non-unitary quench processes. We have further confirmed the decisive role of $\mathcal{PT}$-symmetry on the existence of fixed points and skyrmions in non-unitary dynamics.

Emergent momentum-time skyrmions reported here are intimately connected with the recently observed dynamic quantum phase transitions in similar systems~\cite{pxdqpt,wytheory18}. In fact, both emergent topological phenomena necessarily exist in the presence of fixed points of different kinds, for both unitary and non-unitary dynamics. With the highly flexible control of photonic QW protocols, it would be interesting to investigate dynamic topological phenomena in higher dimensions or associated with other topological classifications in the future~\cite{poprb}.
Our work thus paves the way for a systematic experimental study of dynamic topological phenomena in both unitary and non-unitary dynamics.

\acknowledgements
This work has been supported by the Natural Science Foundation of China (Grant Nos. 11474049, 11674056, and 11522545) and the Natural Science Foundation of Jiangsu Province (Grant No. BK20160024). WY acknowledges support from the National Key R\&D Program (Grant Nos. 2016YFA0301700,2017YFA0304100). KKW and XZQ contributed equally to this work.

{\it Note added.} During the preparation of this manuscript, we learned a related experiment on emergent momentum-time skyrmions in unitary quench dynamics by X.-Y. Guo et al. (arXiv:1806.09269), where the quench dynamics is simulated using superconducting qubits.

\clearpage
\begin{widetext}
\appendix



\section{$\mathcal{PT}$-symmetric non-unitary QW}

The non-unitary Floquet operator $U$ in Eq.~(\ref{eq:defU}) has passive $\mathcal{PT}$ symmetry, from which we can define $\tilde{U}=\gamma U$ with $\gamma=(1-p)^{-\frac{1}{4}}$. $\tilde{U}$ has active $\mathcal{PT}$ symmetry, with the symmetry operator $\mathcal{PT}=\sum_x|-x\rangle\langle x|\otimes\sigma_3\mathcal{K}$, where $\mathcal{K}$ is the complex conjugation.
As homogeneous QWs have lattice translational symmetry, we write $\tilde{U}$ in momentum space
\begin{align}
\tilde{U}_k =&d_0\sigma_0-id_1\sigma_1-id_2\sigma_2-id_3\sigma_3,\nonumber\\
d_0 =&\alpha\left[\cos(2k)\cos\theta_1\cos\theta_2-\sin\theta_1\sin\theta_2\right],\nonumber\\
d_1 =&i\beta,\\
d_2 =&\alpha\left[\cos(2k)\cos\theta_2\sin\theta_1+\cos\theta_1\sin\theta_2\right],\nonumber\\
d_3 =&-\alpha\sin(2k)\cos\theta_2,\nonumber
\label{eqn:U_nonunitary}
\end{align}
where $\alpha=\frac{\gamma}{2}(1+\sqrt{1-p}), \beta=\frac{\gamma}{2}(1-\sqrt{1-p})$.

Eigenvalues of $\tilde{U}_k$ are $\lambda_{k,\pm}=d_0\mp i\sqrt{1-d_0^2}$, and the corresponding quasienergy $\epsilon_{k,\pm}=i\ln(\lambda_{k,\pm})$. When $d_0^2<1$ for all $k$, the quasienergy is real, and the system is in the $\mathcal{PT}$-symmetry-unbroken regime. Whereas if $d_0^2\geq1$ for some $k$, the $\mathcal{PT}$ symmetry is spontaneously broken and the quasienergy is imaginary in the corresponding momentum range.

\section{Winding numbers of non-unitary QWs}

Non-unitary QWs governed by $U$ possess topological properties, which are characterized by winding numbers defined through the global Berry phase
$\nu=\varphi_\text{B}/2\pi$. Here, $\varphi_\text{B}=\varphi_{Z+}+\varphi_{Z-}$, with the generalized Zak phases
\begin{align}
\varphi_{Z\pm} =-i\oint \text{d}k\frac{\langle\chi_{k,\pm}|\frac{\text{d}}{\text{d}k}|\psi_{k,\pm}\rangle}{\langle\chi_{k,\pm}|\psi_{k,\pm}\rangle}.
\end{align}
The integral above is over the first Brillioun zone and $\langle\chi_{k,\mu}|$ and $|\psi_{k,\mu}\rangle$ ($\mu=\pm$) are respectively the left and right eigenstates of $U_k$, defined through $U^{\dag}_k|\chi_{k,\mu}\rangle=\lambda_\mu^* |\chi_{k,\mu}\rangle$ and $U_k|\psi_{k,\mu}\rangle=\lambda_\mu|\psi_{k,\mu}\rangle$, respectively.

\section{Fixed points and dynamical Chern numbers in QW dynamics}

Non-unitary time evolution of the system is captured by the non-Hermitian density matrix
\begin{align}
\rho(k,t):=\frac{|\psi_k(t)\rangle\langle\chi_k(t)|}{\langle\chi_k(t)|\psi_k(t)\rangle}=\frac{1}{2}\left[\tau_0+\bm{n}(k,t)\cdot\bm{\tau}\right],
\label{eqn:rhomain}
\end{align}
where the time-evolved state $|\psi_k(t)\rangle=\sum_{\mu=\pm} c_\mu e^{-i \epsilon^\text{f}_{k,\mu}t}|\psi^{\text{f}}_\mu\rangle$, the associated state $\langle\chi_k(t)|:=\sum_\mu c^*_\mu e^{i \epsilon^{\text{f}*}_{k,\,u}t}\langle\chi^{\text{f}}_{k,\mu}|$, $c_\mu=\langle\chi^{\text{f}}_{k,\mu}|\psi^{\text{i}}_{k,-}\rangle$, and $\langle\chi^{\text{f}}_{k,\mu}|$ $\left(|\psi^{\text{f}}_{k,\mu}\rangle\right)$ is the left (right) eigenvector of $U^\text{f}_k$, with the biorthonormal conditions $\langle\chi^{\text{f}}_{k,\mu}|\psi^{\text{f}}_{k,\nu}\rangle=\delta_{\mu\nu}$ and $\sum_\mu|\psi^{\text{f}}_{k,\mu}\rangle\langle\chi^{\text{f}}_{k,\mu}|=1$.
Under the definition (\ref{eqn:rhomain}), the expectation value of any observable $A_k$ at time $t$ is $\text{Tr} \left[\rho(k,t) A_k\right]=\langle \chi_k(t)|A_k|\psi_k(t)\rangle$.
We also note that $\bm{n}(k,t)=\text{Tr}\left[\rho(k,t)\cdot\bm{\tau}\right]$, with $\{\tau_i\}$ satisfying the standard $\mathfrak{su}(2)$ commutation relations.


Following the convention of the main text, we denote the final Flouqet operator in each quasimomentum~$k$-sector as $U^\text{f}_k$ and the corresponding quasienergy as $\pm E^\text{f}_k$. When $E_k^\text{f}$ is real, we have
\begin{equation}
\begin{split}
n_0&=c^*_+ c_+ + c^*_- c_-,\\
n_1&=\frac{1}{n_0}(c^*_-c_+e^{-i2E_k^{\text{f}}t}+\text{c.c.}),\\
n_2&= \frac{i}{n_0}(c^*_-c_+e^{-i2E_k^{\text{f}}t}-\text{c.c.}),\\
n_3&=\frac{1}{n_0}(c^*_+ c_+ - c^*_- c_-).
\end{split}
\label{Eq:cpmR}
\end{equation}
By contrast, when $E^{\text{f}}_k$ is imaginary, and assuming $\text{Im}(E^{\text{f}}_k)>0$, we have
\begin{equation}
\begin{split}
n_0&=c^*_+ c_+e^{-i2E^\text{f}_kt} + c^*_- c_-e^{i2E^\text{f}_kt},\\
n_1&=\frac{1}{n_0}(c^*_-c_++c.c.),\\
n_2&= \frac{i}{n_0}(c^*_-c_+-c.c.),\\
n_3&= \frac{1}{n_0}(c^*_+ c_+e^{-i2E^\text{f}_kt} - c^*_- c_-e^{i2E^{\text{f}}_kt}).
\end{split}
\label{Eq:cpmC}
\end{equation}

From these expressions, it is straightforward to visualize dynamics of $\bm{n}(k,t)$ on a Bloch sphere as illustrated in Fig.~\ref{fig:bloch}(b) and discussed in the main text. In particular, when $U^\text{f}$ is in the $\mathcal{PT}$-symmetry-unbroken regime, fixed points occur at momenta with $c_-=0$ or $c_+=0$, which we identify as two different types of fixed points.

\section{Dynamic Chern number}

When $U_\text{f}$ is in the $\mathcal{PT}$-symmetry-unbroken regime, periodical evolution of the density matrix in each $k$-sector gives rise to a temporal $S^1$ topology.
In the presence of fixed points, each momentum submanifold between two adjacent fixed points can be combined with the $S^1$ topology in time to form a momentum-time submanifold $S^2$, which can be mapped to the Bloch sphere associated with the vector $\bm{n}(k,t)$.
These $S^2\rightarrow S^2$ mappings define a series of dynamic Chern numbers
\begin{align}
C_{mn} = \frac{1}{4\pi}\int_{k_m}^{k_{n}}\text{d}k\int_{0}^{t_0}\text{d}t\left[\bm{n}(k,t)\times\partial_t \bm{n}(k,t)\right]\cdot\partial_k \bm{n}(k,t),
\label{eqn:chern}
\end{align}
where $k_m$ and $k_n$ denote two neighboring fixed points, and $t_0=\pi/E^\text{f}_{k}$.
For quenches between Hamiltonians with different winding numbers, the dynamic Chern numbers are quantized, with values dependent on the nature of fixed points at $k_m$ and $k_{n}$: $C_{mn}=1$ when $c_+(k_{m})=0$ and $c_-(k_{n})=0$; $C_{mn}=-1$ when $c_-(k_{m})=0$ and $c_+(k_{n})=0$. When the two fixed points are of the same kind, $C_{mn}=0$.

According to its definition in Eq.~(\ref{eqn:chern}), a finite Chern number in a momentum-time submanifold corresponds to the emergence of momentum-time skyrmions in the same submanifold.

\section{Constructing density matrix from direct measurements}

The non-Hermitian density matrix $\rho(k,t)$ is related to the Hermitian one $\rho'(k,t):=|\psi_k(t)\rangle\langle\psi_k(t)|$ through
\begin{align}
\rho(k,t)=\frac{\rho'(k,t)\cdot\sum\limits_{\mu=\pm}\ket{\chi_{k,\mu}^\text{f}}\bra{\chi_{k,\mu}^\text{f}}}{\text{Tr}\Big[\rho^\prime(k,t)\cdot\sum\limits_{\mu=\pm}\ket{\chi_{k,\mu}^\text{f}}\bra{\chi_{k,\mu}^\text{f}}\Big]},
\label{eqn:rhorelate}
\end{align}
where we have used the biorthonormal conditions $\langle\chi^{\text{f}}_{k,\mu}|\psi^{\text{f}}_{k,\nu}\rangle=\delta_{\mu\nu}$ and $\sum_{\mu=\pm}|\psi^{\text{f}}_{k,\mu}\rangle\langle\chi^{\text{f}}_{k,\mu}|=1$.

We then experimentally measure $\rho'(k,t)$ and construct $\rho(k,t)$ and $\bm{n}(k,t)$ using Eq.~(\ref{eqn:rhorelate}).
More specifically, we have
\begin{align}
\rho'(k,t)&=\ket{\psi_k(t)}\bra{\psi_k(t)}\nonumber\\
&=\frac{1}{2}\sum_{j=0}^{3}\text{  }\sum_{x_1,x_2} \text{e}^{-ik(x_1-x_2)}\bra{\psi_{x_2}(t)}\sigma_j\ket{\psi_{x_1}(t)}\sigma_j,
\end{align}
where $|\psi_x(t)\rangle$ is the coin state on site $x$ at the $t$-the time step.
And we experimentally measure $\bra{\psi_{x_2}(t)}\sigma_j\ket{\psi_{x_1}(t)}$ ($j=0,1,2,3$) for each pair of positions $x_1$ and $x_2$ directly. In the case of $x_1=x_2$, we perform projective measurements on the polarizations of photons at each position. In the case of $x_1\neq x_2$, we employ interference-based measurements to construct the the matrix element $\bra{\psi_{x_2}(t)}\sigma_j\ket{\psi_{x_1}(t)}$ from experimental data.

\section{Experimental details}

Experimentally, we implement the coin operator $R(\theta)=\one_\text{w}\otimes \text{e}^{-i\theta\sigma_2}$, the shift operator $S=\sum_x\left(\ket{x-1}\bra{x}\otimes\ket{H}\bra{H}+\ket{x+1}\bra{x}\otimes\ket{V}\bra{V}\right)$, and the loss operator
$M=\one_\text{w}\otimes\left(\ket{+}\bra{+}+\sqrt{1-p}\ket{-}\bra{-}\right)$, following the approach outlined in Ref.~\cite{pxdqpt}. Here,  $\ket{\pm}=(\ket{H}\pm\ket{V})/\sqrt{2}$, $\sigma_2=i(-\ket{H}\bra{V}+\ket{V}\bra{H})$ is the standard Pauli operator under the polarization basis, $\ket{x}$ ($x\in L$) denotes the spatial mode, $\one_\text{w}=\sum_x\ket{x}\bra{x}$, and the loss parameter $p=0.36$ for non-unitary QWs in our experiment.

Now we detail the experimental detection of matrix elements $\bra{\psi_{x_2}(t)}\sigma_i\ket{\psi_{x_1}(t)}$ ($i=0,1,2,3$), which are critical for the detection of momentum-time skyrmions.

For matrix elements with $x_1=x_2=x$, we perform polarization analysis on each lattice site in the basis $\{\ket{H},\ket{V},\ket{L}=(\ket{H}-i\ket{V})/\sqrt{2},\ket{D}=(\ket{H}+\ket{V})/\sqrt{2}\}$. Denoting probabilities of photons measured in the four basis states respectively as $P_{\text{H}}(x,t)$, $P_{\text{V}}(x,t)$, $P_{\text{L}}(x,t)$ and $P_{\text{D}}(x,t)$, we have the matrix elements
\begin{align}
&\bra{\psi_x(t)}\sigma_0\ket{\psi_x(t)}=P_\text{H}(x,t)+P_\text{V}(x,t),\nonumber\\
&\bra{\psi_x(t)}\sigma_1\ket{\psi_x(t)}=2P_\text{D}(x,t)-P_\text{H}(x,t)-P_\text{V}(x,t)\nonumber\\
&\bra{\psi_x(t)}\sigma_2\ket{\psi_x(t)}=-2P_\text{L}(x,t)+P_\text{H}(x,t)+P_\text{V}(x,t)\nonumber\\
&\bra{\psi_x(t)}\sigma_3\ket{\psi_x(t)}=P_\text{H}(x,t)-P_\text{V}(x,t).\nonumber
\end{align}

For matrix elements with $x_1\neq x_2$, we need to measure
\begin{align}
&\bra{\psi_{x_2}(t)}\sigma_0\ket{\psi_{x_1}(t)}=a^*_{x_2}(t)a_{x_1}(t)+b^*_{x_2}(t)b_{x_1}(t),\nonumber\\
&\bra{\psi_{x_2}(t)}\sigma_1\ket{\psi_{x_1}(t)}=a^*_{x_2}(t)b_{x_1}(t)+b^*_{x_2}(t)a_{x_1}(t)\nonumber\\
&\bra{\psi_{x_2}(t)}\sigma_2\ket{\psi_{x_1}(t)}=-ia^*_{x_2}(t)b_{x_1}(t)+ib^*_{x_2}(t)a_{x_1}(t)\nonumber\\
&\bra{\psi_{x_2}(t)}\sigma_3\ket{\psi_{x_1}(t)}=a^*_{x_2}(t)a_{x_1}(t)-b^*_{x_2}(t)b_{x_1}(t),\nonumber
\end{align}
where we have denoted $\ket{\psi_{x}(t)}=\big[a_{x}(t),b_{x}(t)\big]^\text{T}$.
Here, instead of projective measurements, we perform interference-based measurements. As illustrated in Fig.~1 of the main text, photons in spatial modes $x_1$ and $x_2$ are injected into the same spatial mode by passing through HWPs (H$_1$ and H$_2$) and BDs.
After passing through H$_1$ and H$_2$ with specific setting angles, the polarization states of the photons are prepared into one of the following four states:
\begin{align}
&\ket{\phi_1}_\text{c}\propto\Big[a_{x_1}(t),a_{x_2}(t)\Big]^\text{T} \text{  when }\text{H}_1\text{ at }0,\text{H}_2\text{ at }45^\circ,\nonumber \\
&\ket{\phi_2}_\text{c}\propto\Big[b_{x_1}(t),-b_{x_2}(t)\Big]^\text{T} \text{  when }\text{H}_1\text{ at }45^\circ,\text{H}_2\text{ at }0 \nonumber\\ &\ket{\phi_3}_\text{c}\propto\Big[b_{x_1}(t),a_{x_2}(t)\Big]^\text{T} \text{  when }\text{H}_1\text{ at }45^\circ,\text{H}_2\text{ at }45^\circ,  \nonumber\\
&\ket{\phi_4}_\text{c}\propto\Big[a_{x_1}(t),b_{x_2}(t)\Big]^\text{T} \text{  when }\text{removing H$_1$ and H$_2$}.\nonumber
\end{align}
We then apply a projective measurement $\{\ket{L}\bra{L},\ket{D}\bra{D}\}$ with a QWP, an HWP and a PBS to obtain probabilities of photons in the basis states $\{\ket{L},\ket{D}\}$. Depending on their polarization states $|\phi_j\rangle_\text{c}$ ($j=1,2,3,4$) prior to the projective measurement, we denote the measured probabilities as $P^j_{L}(x_1,x_2,t)$ and $P^j_{D}(x_1,x_2,t)$, respectively.

We are then able to calculate both the real and imaginary parts of $\bra{\psi_{x_2}(t)}\sigma_i\ket{\psi_{x_1}(t)}$ ($i=0,1,2,3$) through
\begin{align}
&\text{Re}\big[\bra{\psi_{x_2}(t)}\sigma_0\ket{\psi_{x_1}(t)}\big]=P^1_\text{D}(x_1,x_2,t)-P^2_\text{D}(x_1,x_2,t)-\frac{P_\text{H}(x_1,t)+P_\text{H}(x_2,t)-P_\text{V}(x_1,t)-P_\text{V}(x_2,t)}{2},\nonumber\\
&\text{Im}\big[\bra{\psi_{x_2}(t)}\sigma_0\ket{\psi_{x_1}(t)}\big]=P^1_\text{L}(x_1,x_2,t)-P^2_\text{L}(x_1,x_2,t)-\frac{P_\text{H}(x_1,t)+P_\text{H}(x_2,t)-P_\text{V}(x_1,t)-P_\text{V}(x_2,t)}{2},\nonumber\\
&\text{Re}\big[\bra{\psi_{x_2}(t)}\sigma_1\ket{\psi_{x_1}(t)}\big]=P^3_\text{D}(x_1,x_2,t)+P^4_\text{D}(x_1,x_2,t)-\frac{P_\text{V}(x_1,t)+P_\text{H}(x_2,t)+P_\text{H}(x_1,t)+P_\text{V}(x_2,t)}{2},\nonumber\\
&\text{Im}\big[\bra{\psi_{x_2}(t)}\sigma_1\ket{\psi_{x_1}(t)}\big]=P^3_\text{L}(x_1,x_2,t)+P^4_\text{L}(x_1,x_2,t)-\frac{P_\text{V}(x_1,t)+P_\text{H}(x_2,t)+P_\text{H}(x_1,t)+P_\text{V}(x_2,t)}{2},\nonumber\\
&\text{Re}\big[\bra{\psi_{x_2}(t)}\sigma_2\ket{\psi_{x_1}(t)}\big]=P^3_\text{L}(x_1,x_2,t)-P^4_\text{L}(x_1,x_2,t)-\frac{P_\text{V}(x_1,t)+P_\text{H}(x_2,t)-P_\text{H}(x_1,t)-P_\text{V}(x_2,t)}{2},\nonumber\\
&\text{Im}\big[\bra{\psi_{x_2}(t)}\sigma_2\ket{\psi_{x_1}(t)}\big]=P^4_\text{D}(x_1,x_2,t)-P^3_\text{D}(x_1,x_2,t)+\frac{P_\text{V}(x_1,t)+P_\text{H}(x_2,t)-P_\text{H}(x_1,t)-P_\text{V}(x_2,t)}{2},\nonumber\\
&\text{Re}\big[\bra{\psi_{x_2}(t)}\sigma_3\ket{\psi_{x_1}(t)}\big]=P^1_\text{D}(x_1,x_2,t)+P^2_\text{D}(x_1,x_2,t)-\frac{P_\text{H}(x_1,t)+P_\text{H}(x_2,t)+P_\text{V}(x_1,t)+P_\text{V}(x_2,t)}{2},\nonumber\\
&\text{Im}\big[\bra{\psi_{x_2}(t)}\sigma_3\ket{\psi_{x_1}(t)}\big]=P^1_\text{L}(x_1,x_2,t)+P^2_\text{L}(x_1,x_2,t)-\frac{P_\text{H}(x_1,t)+P_\text{H}(x_2,t)+P_\text{V}(x_1,t)+P_\text{V}(x_2,t)}{2},\nonumber
\end{align}
from which we construct the corresponding matrix elements.

\section{Quench between FTPs with the same winding number}
\begin{figure*}
\includegraphics[width=\textwidth]{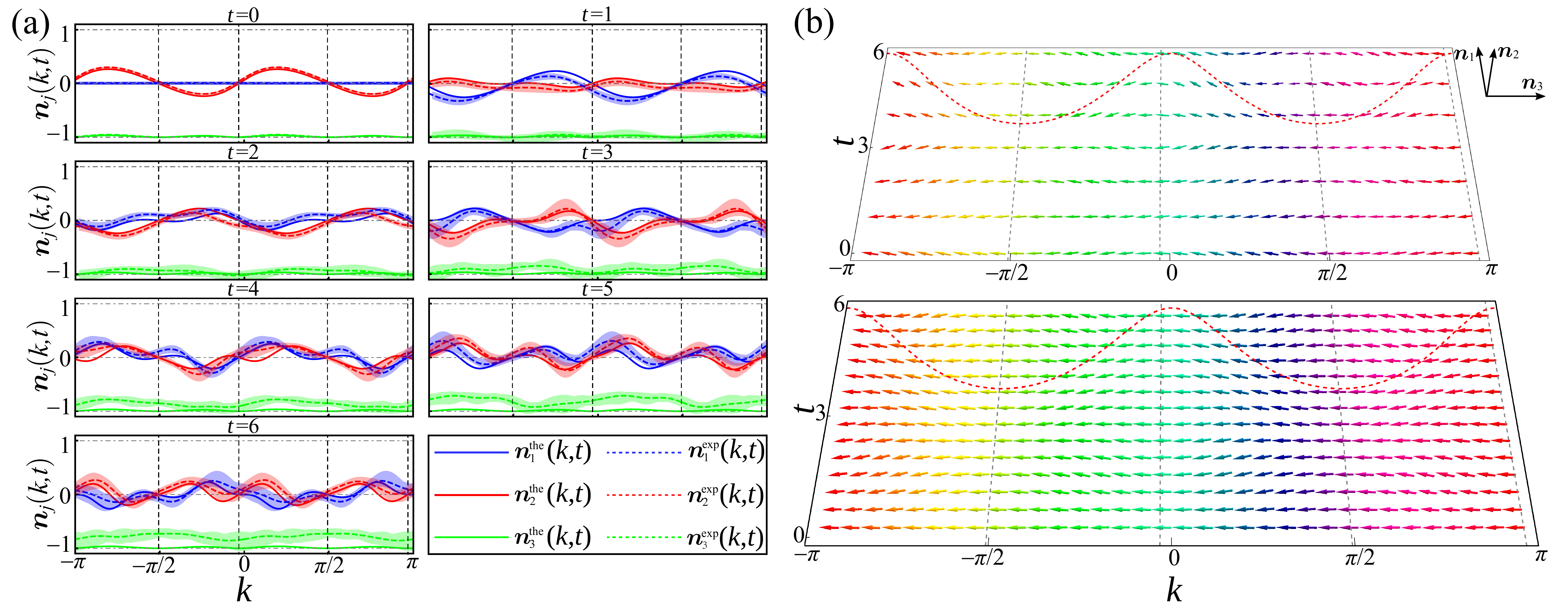}
\caption{Experimental results for the $\mathcal{PT}$-symmetry-unbroken QW dynamics with initial and final coin parameters chosen in the same topological phase. (a) Time-evolution and (b) spin textures of $\bm{n}(k,t)$ in the momentum-time space for a quench process between the initial non-unitary Floquet operator given by $(\theta^\text{i}_1=\pi/4,\theta^\text{i}_2=-\pi/2)$ and the final Floquet operator given by $(\theta^\text{f}_1=7\pi/25,\theta^\text{f}_2=-9\pi/20)$ [black square in Fig.~2(a) of the main text].
Fixed points are located at $\{-1.0319\pi,-0.5069\pi,-0.0319\pi,0.4913\pi\}$.
Red dashed lines in (b) show the momentum-dependent period $\pi/E^\text{f}_k$ for the oscillations of $\bm{n}(k,t)$, which, together with fixed points in momentum space, mark the boundary of the momentum-time submanifolds.
}
\label{fig:trivial}
\end{figure*}

As we have discussed in the main text, when the system is quenched between FTPs with the same winding number, skyrmion-lattice structures are no longer present. This is shown in Fig.~\ref{fig:trivial}, where we set the coin parameters for $U^\text{f}$ as $(\theta^\text{f}_1=7\pi/25,\theta^\text{f}_2=-9\pi/20)$. The initial state and the loss parameter $p$ are the same as those in Fig.~3(b) of the main text. Therefore, the post-quench FTP is in the $\mathcal{PT}$-symmetry unbroken regime with $\nu^\text{f}=0$ and belongs to the same phase regime for $U^\text{i}$. As shown in Fig.~\ref{fig:trivial}(a), dynamics of $\bm{n}(k,t)$ is still oscillatory, however, skyrmion-lattice structures are no longer present in Fig.~\ref{fig:trivial}(b). Note that as $\theta_{1,2}^\text{f}$ are not chosen along the purple dashed lines in Fig.~2(a), the corresponding quasienergy band $E^\text{f}_k$ is not flat, which leads to oscillations of $\bm{n}(k,t)$ with momentum-dependent periods. This is shown in Fig.~\ref{fig:trivial}(b).

\end{widetext}

\end{document}